\documentclass[journal=jacsat,manuscript=reprint,layout=onecolumn]{achemso}
\usepackage{graphicx}
\usepackage{dcolumn}
\usepackage{bm}
\usepackage{amsfonts}
\usepackage{amsmath}
\usepackage{color}
\usepackage{amssymb}
\usepackage{multirow}
\usepackage{epstopdf}
\usepackage{natbib}
\usepackage{epstopdf}
\usepackage{fancyhdr}
\usepackage[utf8]{inputenc}
\usepackage[T1]{fontenc}
\usepackage{mathptmx}
\usepackage{csquotes}

\author{Yingyue Hong}
\affiliation{State Key Laboratory of Molecular Reaction Dynamics, Dalian Institute of Chemical Physics, Chinese Academy of Science, Dalian 116023, China}
\alsoaffiliation{School of Chemical Sciences, University of Chinese Academy of Sciences, Beijing 100049, China}
\author{Jiayu Huang}
\email{jyhuang@dlut.edu.cn}
\affiliation{Department of Physics, Dalian University of Technology, Dalian 116024, China}
\alsoaffiliation{State Key Laboratory of Molecular Reaction Dynamics, Dalian Institute of Chemical Physics, Chinese Academy of Science, Dalian 116023, China}
\author{Dong H. Zhang}%
\email{zhangdh@dicp.ac.cn}
\affiliation{State Key Laboratory of Molecular Reaction Dynamics, Dalian Institute of Chemical Physics, Chinese Academy of Science, Dalian 116023, China}
\alsoaffiliation{School of Chemical Sciences, University of Chinese Academy of Sciences, Beijing 100049, China}

\title{Highly Accurate Description of Long-Range Interactions through the Combination of Neural Networks and Physical Models}

\begin{document}

\begin{tocentry}
    \includegraphics[width=6cm]{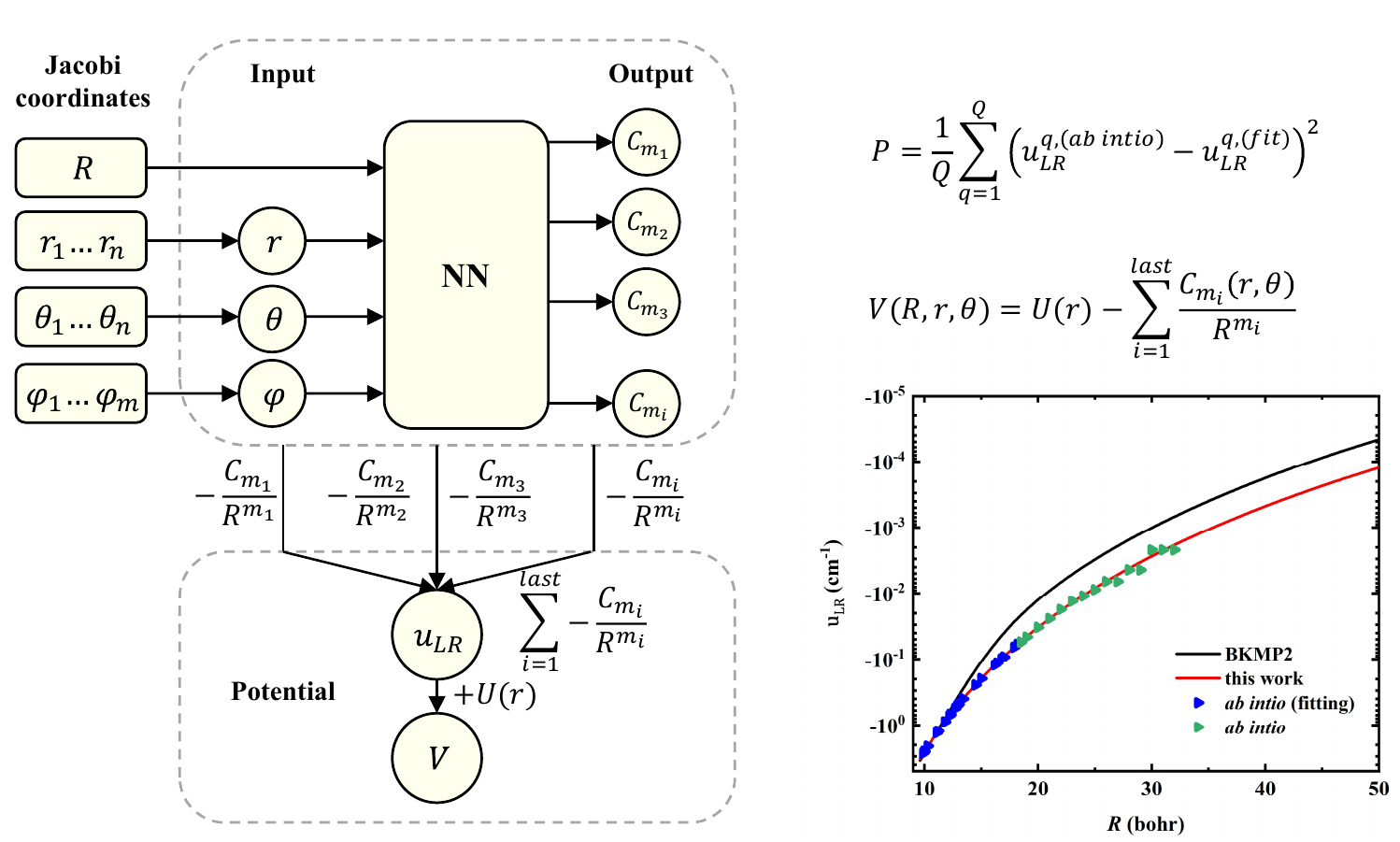}
\end{tocentry}

\begin{abstract}
We present a simple and general way to accurately describe long-range interactions between atoms and molecules through combining neural networks with physical models. Demonstrations on the H$_3$, Li$_3$ and 2KRb systems illustrate the exceptional extrapolation capabilities of the trained model, supported by underlying physical models. More importantly, the model exhibits high accuracy at energy scales below a few hundred millikelvin, where the reliability of $ab~initio$ methods diminishes. 
\end{abstract}

\maketitle 
Understanding and manipulating ultracold collisions ($T<10^{-3}$ K) between atoms and molecules has become a forefront in the fields of physics and chemistry, with implications ranging from quantum information processing\cite{Kaufman2021,PhysRevX.12.021027,PhysRevLett.129.160501,Hu2021}, precision measurement\cite{Liu2021,Christakis2023}, and quantum control of chemical reactions\cite{D1CS01040A,doi:10.1146/annurev-physchem-090419-043244,doi:10.1126/science.abn8525,Yang2022}.

At ultracold temperatures, particles possess minimal kinetic energy. Due to their slow movement, long-range interactions between these particles become dominant. These interactions, extending over significant distances, play a crucial role in shaping particle behavior and notably influence the outcomes of ultracold collisions. Consequently, creating accurate models to represent long-range interactions is essential for the understanding of quantum dynamics during ultracold collisions. Yet, achieving the necessary precision on a scale as small as several nanokelvins presents challenges for the accuracy of potential energy calculations—a difficulty that current ab initio methods may find challenging to overcome.

An alternative approach is employing physics-motivated analytical models to describe long-range interactions. In these models, the long-range behavior of adiabatic Born-Oppenheimer potentials is commonly expressed as a power series in terms of 1/$R$, where $R$ represents the distance between the interacting particles. Guided by high-resolution spectrum experiments, achieving an accurate representation of long-range interactions can be accomplished through finely tuning parameters within physical models. This method has proven successful in the study of ultracold atoms\cite{RevModPhys.78.483}.
It has also been shown that physical models can be applied for representing long-range interactions between ultracold closed-shell molecules by using the standard long-range expansion\cite{10.1063/5.0014805,doi:10.1021/acs.jpca.1c04506,D3CP01753B}, which includes multipole-multipole electrostatic, dispersion, and induction interactions. However, these analytical models often face challenges when attempting to represent the intricate long-range interactions within ultracold open-shell molecules. 

Recent advancements in artificial neural networks (NNs) introduced a powerful tool for modeling chemical systems. The NN-based potential models can estimate interaction energies with near ab initio accuracy while drastically reducing computational costs by orders of magnitude.\cite{10.1063/1.469597, 10.1063/1.472596, doi:10.1021/jp9105585, C1CP21668F, 10.1063/1.4811109, 10.1063/1.4817187, C5SC03689E,10.1063/1.4961454,doi:10.1021/acs.jpclett.7b00784} However, the incorporation of artificial neural networks faces notable limitations when applied to systems influenced by long-range interactions. Firstly, its capacity to extrapolate beyond available data points is restricted, potentially introducing inaccuracies in describing interactions over very long distances. Secondly, the accuracy of the NN method is limited by its reliance on ab initio points. To overcome these challenges, there is a recent trend in the scientific community to combine neural network methods with physics models\cite{doi:10.1021/acs.jpca.2c06778,D2DD00150K,doi:10.1021/acs.jctc.9b00181,10.1063/5.0083669,doi:10.1021/acs.jctc.2c01049}. 
However, the majority of efforts have been concentrated on enhancing neural network performance by incorporating physical models. The intrinsic challenges of extrapolation and accuracy mentioned above have not been effectively resolved. Consequently, there is currently a shortage of potential energy surface (PES) capable of offering an accurate description of long-range interactions for ultracold collisions.

In this paper, we present a new way to accurately describe long-range interactions between atoms and molecules. The method combines neural networks with physical models, offering a simple and general solution. Demonstrations on the H$_3$, Li$_3$ and 2KRb systems illustrate the exceptional extrapolation capabilities of the trained model, supported by underlying physical models. Moreover, the model exhibits superior accuracy at energy scales below a few hundred millikelvin, where the reliability of ab initio points becomes uncertain. 

We start by considering the Jacobi coordinates for a triatomic system A+BC, denoted as $\left( R, r, \theta \right)$. Here, $R$ represents the distance from A to the center of mass of BC, $r$ is the bond length within BC, and $\theta$ is the angle between $R$ and $r$. According to perturbation theory, the system energy can be expressed 
\begin{equation}
V\left ( R,r ,\theta \right ) = U\left ( r \right ) +u_{LR}\left ( R, r ,\theta \right )
\end{equation}
This energy is divided into two components: $U\left ( r \right )$, representing the inner energy of molecule as the distance $R$ goes to infinity, and $u_{LR}\left ( R, r ,\theta \right )$, which characterizes the long-range interaction energy. The long-range interaction energy can be further expanded as a summation of terms with inverse powers of $R$: 
\begin{equation}
u_{LR}\left ( R, r ,\theta \right )=-\sum_{i}^{}\frac{C_{m_{i}}\left (  r ,\theta \right ) }{R^{m_{i}}}  
\label{eq:ulr}
\end{equation}
where $C_{m_{i}}\left (  r ,\theta \right )$ denotes coefficients that depend on $r$ and $\theta$. The values of $m_{i}$ need to be adjusted based on the physical characteristics of specific system.  

The very idea of this study is to introduce neural networks to represent $C_{m_{i}}\left (  r ,\theta \right )$. The advantage of neural networks in accurately approximating non-linear functions provides increased flexibility for handling complex expressions of $C_{m_{i}}\left (  r ,\theta \right )$. To implement this concept, we adopted feed-forward multilayer networks for the construction of $C_{m_{i}}\left (  r ,\theta \right )$. These networks comprise an input layer, several hidden layers, and an output layer. Input data undergoes processing as it forwards through the hidden layers, with each element from the previous layer connecting to every component in the subsequent layer through suitable weight coefficients. The entire process is mathematically described by 

\begin{equation}
n^{m}_{i}=\sum_{j=1}^{S^{m-1}}\left(w^{m}_{i,j}a^{m-1}_{j}\right)+b_{i}^{m}
\label{eq:4}
\end{equation}
\begin{equation}
a^{m}_{i}=f^{m}\left(n^{m}_{i}\right)
\label{eq:5}
\end{equation}
in which $n^{m}_{i}$ stands for the net input of neuron $i$ in the $m$-th layer, $a^{m}_{i}$ represents the output of neuron $i$ in the $m$-th layer activation functions. Commonly used functions to denote activation functions $f^{m}$ include hyperbolic tangent, linear function, and logistic. $w^{m}_{i,j}$ denotes the weight at row $i$ and column $j$ between layer $m$-1 and layer $m$. $b_{i}^{m}$ represents the bias of neuron $i$ in layer $m$.

\begin{figure}[htbp!]
    \includegraphics[width=\columnwidth]{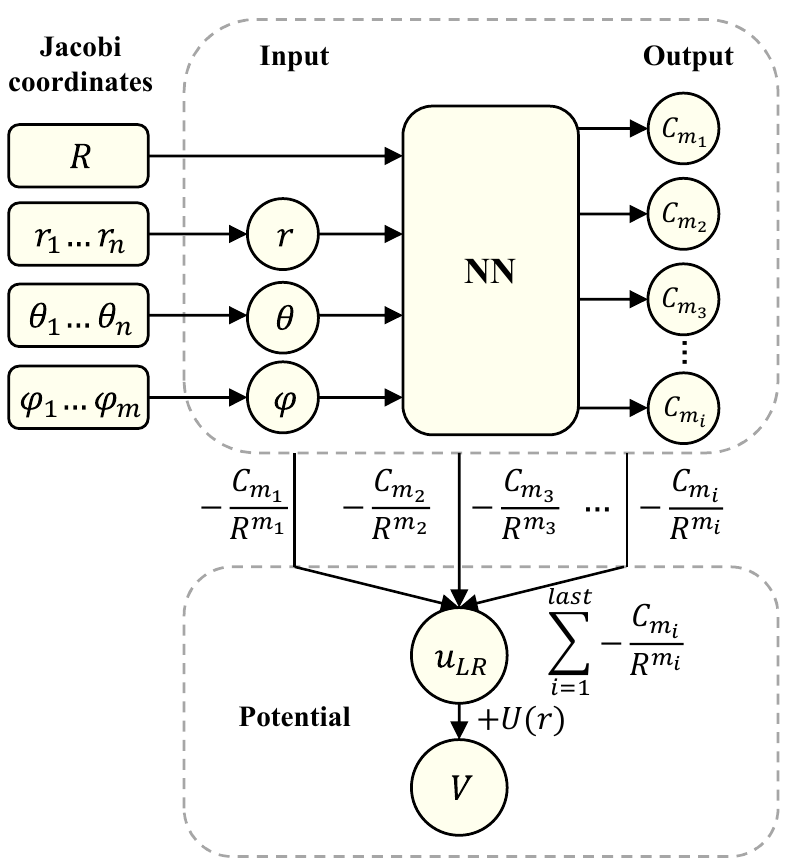}
    \caption{The schematic diagram illustrates the fitting process for long-range intereaction. The initial step involves processing the Jacobi coordinates of the system. $r$, $\theta$ and $\varphi $ are normalized and included as input layers in the neural network. Following the fitting process through the neural network, we obtain values for $C_{m_{i}}$. These parameters are then utilized to calculate $u_{LR}$. With these acquired values, the potential energy in the long-range region is determined by adding the inner energy $U\left ( r_i \right )$ of molecules to $u_{LR}$. }
    \label{NN_illustration} 
\end{figure}

The illustration in Figure \ref{NN_illustration} shows the neural network structure. The input layer of the neural network consists of $\vec{r}_{i}$, $\vec{\theta}_{i}$, and $\vec{\varphi}_{j}$, while the output layer is $C_{m{i}}$. Once $C_{m_{i}}$ is obtained,  the calculation of $u_{LR}$ follows using Eq~(\ref{eq:ulr}). Therefore, The formulation of the loss function $P$ can be expressed as: 
\begin{equation}
P=\frac{1}{Q} \sum_{q=1}^{Q}\left( u_{LR}^{q,\left ( ab~intio \right ) } - u_{LR}^{q,\left ( fit \right ) }\right )^{2}  
\end{equation}
The objective is to minimize the loss function through the Levenberg-Marquardt algorithm iteration. Notably, the input layer of neural network does not include $R$ due to the independence of $C_{m_{i}}$ from $R$. Instead, $R$ is passed as a parameter to the internal optimization process of the neural network. Hence, the Jacobian matrix can be expressed as:
\begin{equation}
\frac{\partial e\left ( \vec{w} \right ) }{\partial w_{j}} =-\sum_{i=0}^{last}\frac{ 1}{R^{m_{i}}}\frac{\partial C_{m_{i}}}{\partial w_{j}}
\end{equation}
While we use the 3-atom collision system in Jacobi coordinates as an example, it is important to note that this is a general method that can be extended to any coordinate system and dimension.

\begin{figure}[htbp!]
    \includegraphics[width=\columnwidth]{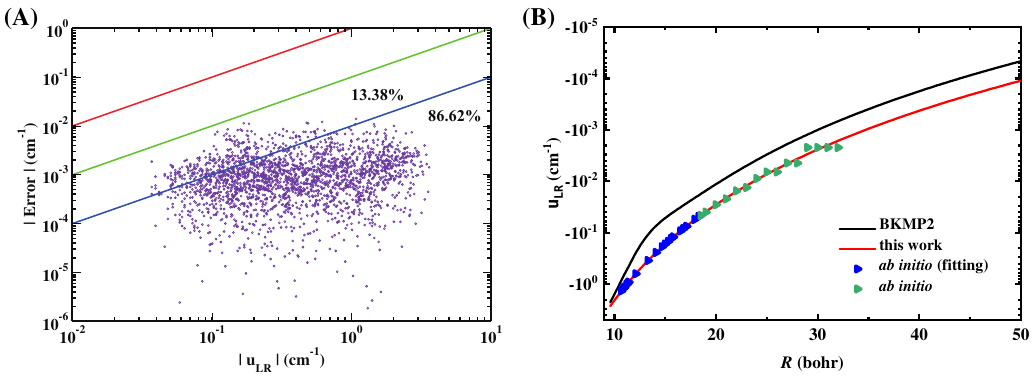}
    \caption{(A) Plot of the distribution of errors in the ground state potential energy surface (PES) for the H$_3$ system. The $ab~initio$ potential energy is plotted on the horizontal axis, while the absolute error is represented on the vertical axis. The red, green, and blue lines divide the points into regions with errors exceeding 100\%, 10\%, and 1\%, respectively. Each region is labeled with the corresponding percentage of error points; (B) One-dimensional plot of $u_{LR}$ energies calculated from fitting results, $ab~initio$ methods, and BKMP2 PES. The plot is generated at $\theta = 90^\circ$ and the equilibrium bond length ($r = 1.4$ Bohr).}
    \label{h3_picture} 
\end{figure}

We applied the new method to the simplest triatomic collision system: H+H$_2$. The system is an ideal choice for testing the accuracy of PES, as $ab~initio$ calculations can provide the most accurate results for it. 
The $ab~initio$ data were calculated through the Molpro program\cite{molpro} by using the UCCSD-T method\cite{DEEGAN1994321,10.1063/1.480886} at the aug-cc-pv5z level\cite{10.1063/1.456153}. 
We perform calculations at 2249 points, ranging from 9.6 to 18.3 bohr for $R$ and 1.0 to 4.0 bohr for $r$. The neural network structure employed was 2-30-3, with the values of $m_{i}$ chosen as 6, 8, and 10 to represent the long-range interaction. The activation functions are logsig and line, and the testing set comprises 20\% of the data. 
To ensure the reliablity of the $ab~initio$ points, most of training points is refined to energies higher than 0.05 cm$^{-1}$. 
The obtained root-mean-square error (RMSE) was 0.00209 cm$^{-1}$. 
Figure \ref{h3_picture}(A) shows the distribution of errors of the fitting results for the long-range potential energy of H$_3$ system. 
It is clear that the errors tend to stay around 10$^{-3}$ cm$^{-1}$, showing that our fitting is quite accurate. All points have a relative error less than 10\%, and a large majority (86.62\%) have a relative error less than 1\%. 
Figure \ref{h3_picture}(B) illustrates the variation of the long-range interaction $u_{LR}$ with the distance between the H atom and H$_2$ molecule at a specific configuration. The energy values are calculated from fitting results, $ab~initio$ methods, and the well-known BKMP2 PES. 
It is evident that the value of $ab~initio$ points become unreliable at energy below 10$^{-2}$ cm$^{-1}$. In contrast, our model shows robust extrapolation capabilities, yielding accurate values over extended distances. 
This demonstrates that the new method overcomes the inherent challenges associated with extrapolation and accuracy in pure neural network methods, and has strong potential to be used for ultracold collisions studies.

\begin{figure}[htbp!]
    \includegraphics[width=\columnwidth]{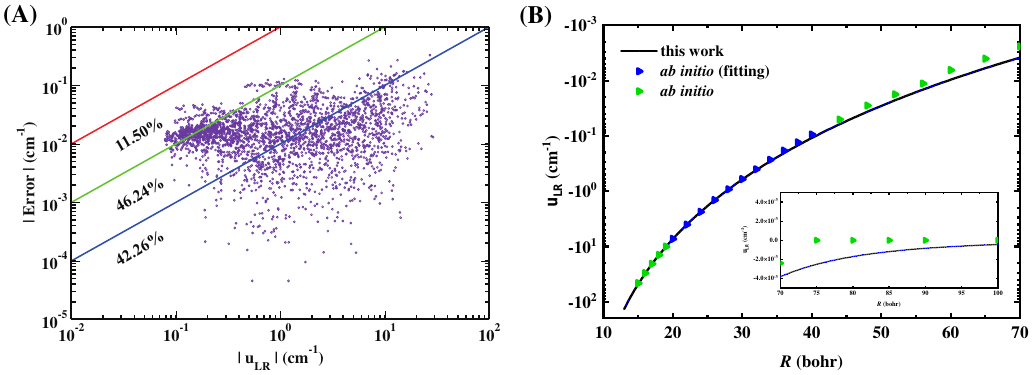}
    \caption{(A) Plot of the distribution of errors in the ground state potential energy surface for the Li$_3$ system, similar to the representation in Figure \ref{h3_picture}(A); (B) One-dimensional plot of $u_{LR}$ energies calculated from  fitting results compared with $ab~initio$ methods. The plot is generated at $\theta = 90^\circ$ and the equilibrium bond length ($r = 5.2$ Bohr). The inset shows the results over extended distance. }
    \label{Li3_picture} 
\end{figure}

Ultracold molecule collision experiments primarily focus on collisions involving alkali metal molecules. To show the capability of our method in describing the long-range interaction of alkali metal molecules systems, we applied it to model the Li+Li$_2$ collision. We employed Molpro program with the UCCSD-T/AVTZ method\cite{DEEGAN1994321,10.1063/1.480886,prascher2011gaussian} to calculate 2669 points, with $R$ ranging from 20 to 40 bohr and $r$ from 4.0 to 9.4 bohr. The neural network structure is 2-50-3, and the output layer consists of C$_6$, C$_8$, and C$_{10}$. The activation functions are logsig and line, and the testing set comprises 10\% of the data. 
The testing set comprises 10\% of the data. Most of training points is refined to energies higher than 0.05 cm$^{-1}$, results in a RMSE of 0.0346 cm$^{-1}$. The comparison between Figure \ref{h3_picture}(B) and Figure \ref{Li3_picture}(B) reveals that the long-range interaction between Li and Li$_2$ is considerably stronger than that between H and H$_2$ at the same distance, showing the importance of accurate long-range PES. 
From Figure \ref{Li3_picture}(A), it is clear that the errors stabilize around 0.01 cm$^{-1}$, with the relative error of fitting results exceeding 10\% at energies below 0.1 cm$^{-1}$. However, we attribute this error to the unreliability of $ab~initio$ calculations rather than our model. This interpretation is supported by Figure \ref{Li3_picture}(B), which indicates that $ab~initio$ calculations become unreliable at energy levels below 0.1 cm$^{-1}$. Therefore, it is reasonable to conclude that the accuracy of the $ab~initio$ method is limited to 0.1 cm$^{-1}$ for the Li$_3$ system. This agrees with recent experimental evidence, indicating challenges even for the gold standard $ab~initio$ method in providing accurate intermolecular potentials at energy levels of several cm$^{-1}$\cite{Vogels2018}. On the contrary, our model demonstrates strong extrapolation capabilities, accurately predicting values over a wide range of intermolecular distances, showing  its reliability for various systems.

\begin{figure}[htbp!]
    \includegraphics[width=\columnwidth]{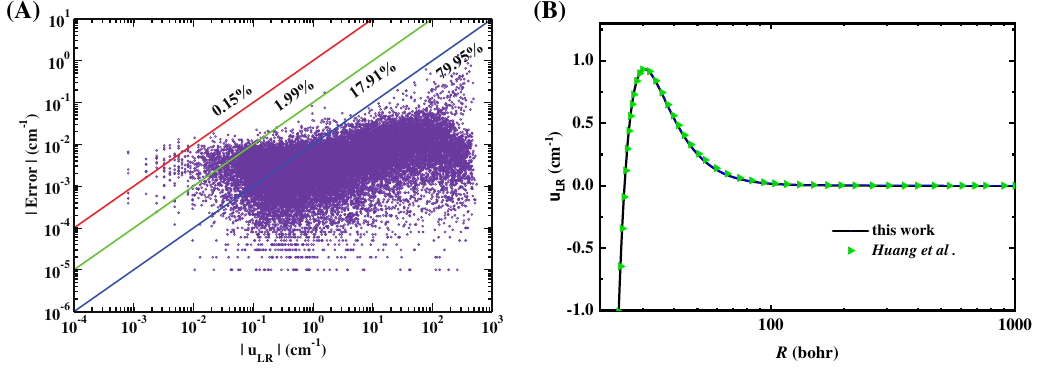}
    \caption{(A) Plot of the distribution of errors in the ground state potential energy surface for the KRb+KRb system, similar to the representation in Figure \ref{h3_picture}(A); (B) One-dimensional plot of $u_{LR}$ energies calculated from fitting results and the analytical PES constructed by Huang and her coworkers. The plot is generated at $\theta_1 = 90^\circ$, $\theta_2 = 90^\circ$, $\phi = 0^\circ$, and the equilibrium bond length ($r_1 = r_2 = 7.66$ Bohr).
    }
    \label{2KRb_picture} 
\end{figure}

To show the applicability of the new method to higher-dimensional systems, we applied it to the 2KRb system, which is the most widely studied ultracold molecule-molecule colllsion system and has can be denoted in the Jacobi coordinates as $\left( R, r_1, r_2, \theta_{1}, \theta_{2}, \phi\right)$. Recently, Huang and her coworkers developed an analytical long-range PES for the long-range interaction of KRb+KRb collisions. \cite{doi:10.1021/acs.jpca.1c04506}
We randomly selected 29,706 points from the newly developed long-range PES, with $R$ ranging from 16 bohr to 64 bohr and two $r$ values ranging from 5.67 to 8.87. 
The neural network structure is 5-400-4, and the output layer consists of C$_3$, C$_4$, C$_5$, and C$_6$, guided by the formula used to construct the analytical PES. The activation functions are logsig and line, and the testing set comprises 10\% of the data. As shown in Figure \ref{2KRb_picture}(A), nearly all points (97.86\%) have a relative error less than 10\%, and a large majority (79.95\%) have a relative error less than 1\%, resulting a RMSE of 0.0244 cm$^{-1}$. More interestingly, Figure \ref{2KRb_picture}(B) illustrates that the trained model accurately predicts results for collision configurations even in the absence of sampling over distances up to 1000 bohr. This demonstrates that our method inherits the strong non-linear fitting capability of NN methods. Therefore, this method is well suitble for higher-dimensional systems. We are confident that the new method has the potential to perform effectively in the investigation of ultracold molecule-molecule collisions, a subject currently of great interest.

To summarize, a simple and general method is developed for describing long-range interactions between atoms and molecules by combining neural networks with physical models. Tests on the H$_3$, Li$_3$, and 2KRb systems demonstrate that this method has strong extrapolation capabilities and accuracy over very long distances. This illustrates that the new method overcomes the inherent challenges associated with extrapolation and accuracy in pure neural network methods, showing its strong potential for use in the study of ultracold collisions.

\clearpage

\clearpage

\clearpage

\clearpage

\clearpage


\begin{mcitethebibliography}{35}
	\providecommand*\natexlab[1]{#1}
	\providecommand*\mciteSetBstSublistMode[1]{}
	\providecommand*\mciteSetBstMaxWidthForm[2]{}
	\providecommand*\mciteBstWouldAddEndPuncttrue
	{\def\EndOfBibitem{\unskip.}}
	\providecommand*\mciteBstWouldAddEndPunctfalse
	{\let\EndOfBibitem\relax}
	\providecommand*\mciteSetBstMidEndSepPunct[3]{}
	\providecommand*\mciteSetBstSublistLabelBeginEnd[3]{}
	\providecommand*\EndOfBibitem{}
	\mciteSetBstSublistMode{f}
	\mciteSetBstMaxWidthForm{subitem}{(\alph{mcitesubitemcount})}
	\mciteSetBstSublistLabelBeginEnd
	{\mcitemaxwidthsubitemform\space}
	{\relax}
	{\relax}
	
	\bibitem[Kaufman and Ni(2021)Kaufman, and Ni]{Kaufman2021}
	Kaufman,~A.~M.; Ni,~K.~K. {Quantum science with optical tweezer arrays of
		ultracold atoms and molecules}. \emph{Nat. Phys.} \textbf{2021}, \emph{17},
	1324--1333\relax
	\mciteBstWouldAddEndPuncttrue
	\mciteSetBstMidEndSepPunct{\mcitedefaultmidpunct}
	{\mcitedefaultendpunct}{\mcitedefaultseppunct}\relax
	\EndOfBibitem
	\bibitem[Jenkins \latin{et~al.}(2022)Jenkins, Lis, Senoo, McGrew, and
	Kaufman]{PhysRevX.12.021027}
	Jenkins,~A.; Lis,~J.~W.; Senoo,~A.; McGrew,~W.~F.; Kaufman,~A.~M. Ytterbium
	Nuclear-Spin Qubits in an Optical Tweezer Array. \emph{Phys. Rev. X}
	\textbf{2022}, \emph{12}, 021027\relax
	\mciteBstWouldAddEndPuncttrue
	\mciteSetBstMidEndSepPunct{\mcitedefaultmidpunct}
	{\mcitedefaultendpunct}{\mcitedefaultseppunct}\relax
	\EndOfBibitem
	\bibitem[Gonz\'alez-Cuadra \latin{et~al.}(2022)Gonz\'alez-Cuadra, Zache,
	Carrasco, Kraus, and Zoller]{PhysRevLett.129.160501}
	Gonz\'alez-Cuadra,~D.; Zache,~T.~V.; Carrasco,~J.; Kraus,~B.; Zoller,~P.
	Hardware Efficient Quantum Simulation of Non-Abelian Gauge Theories with
	Qudits on Rydberg Platforms. \emph{Phys. Rev. Lett.} \textbf{2022},
	\emph{129}, 160501\relax
	\mciteBstWouldAddEndPuncttrue
	\mciteSetBstMidEndSepPunct{\mcitedefaultmidpunct}
	{\mcitedefaultendpunct}{\mcitedefaultseppunct}\relax
	\EndOfBibitem
	\bibitem[Hu \latin{et~al.}(2021)Hu, Liu, Nichols, Zhu, Qu{\'{e}}m{\'{e}}ner,
	Dulieu, and Ni]{Hu2021}
	Hu,~M.~G.; Liu,~Y.; Nichols,~M.~A.; Zhu,~L.; Qu{\'{e}}m{\'{e}}ner,~G.;
	Dulieu,~O.; Ni,~K.~K. {Nuclear spin conservation enables state-to-state
		control of ultracold molecular reactions}. \emph{Nat. Chem.} \textbf{2021},
	\emph{13}, 435--440\relax
	\mciteBstWouldAddEndPuncttrue
	\mciteSetBstMidEndSepPunct{\mcitedefaultmidpunct}
	{\mcitedefaultendpunct}{\mcitedefaultseppunct}\relax
	\EndOfBibitem
	\bibitem[Liu \latin{et~al.}(2021)Liu, Hu, Nichols, Yang, Xie, Guo, and
	Ni]{Liu2021}
	Liu,~Y.; Hu,~M.~G.; Nichols,~M.~A.; Yang,~D.; Xie,~D.; Guo,~H.; Ni,~K.~K.
	{Precision test of statistical dynamics with state-to-state ultracold
		chemistry}. \emph{Nature} \textbf{2021}, \emph{593}, 379--384\relax
	\mciteBstWouldAddEndPuncttrue
	\mciteSetBstMidEndSepPunct{\mcitedefaultmidpunct}
	{\mcitedefaultendpunct}{\mcitedefaultseppunct}\relax
	\EndOfBibitem
	\bibitem[Christakis \latin{et~al.}(2023)Christakis, Rosenberg, Raj, Chi,
	Morningstar, Huse, Yan, and Bakr]{Christakis2023}
	Christakis,~L.; Rosenberg,~J.~S.; Raj,~R.; Chi,~S.; Morningstar,~A.;
	Huse,~D.~A.; Yan,~Z.~Z.; Bakr,~W.~S. {Probing site-resolved correlations in a
		spin system of ultracold molecules}. \emph{Nature} \textbf{2023}, \emph{614},
	64--69\relax
	\mciteBstWouldAddEndPuncttrue
	\mciteSetBstMidEndSepPunct{\mcitedefaultmidpunct}
	{\mcitedefaultendpunct}{\mcitedefaultseppunct}\relax
	\EndOfBibitem
	\bibitem[Zhao and Pan(2022)Zhao, and Pan]{D1CS01040A}
	Zhao,~B.; Pan,~J.-W. Quantum control of reactions and collisions at ultralow
	temperatures. \emph{Chem. Soc. Rev.} \textbf{2022}, \emph{51},
	1685--1701\relax
	\mciteBstWouldAddEndPuncttrue
	\mciteSetBstMidEndSepPunct{\mcitedefaultmidpunct}
	{\mcitedefaultendpunct}{\mcitedefaultseppunct}\relax
	\EndOfBibitem
	\bibitem[Liu and Ni(2022)Liu, and
	Ni]{doi:10.1146/annurev-physchem-090419-043244}
	Liu,~Y.; Ni,~K.-K. Bimolecular Chemistry in the Ultracold Regime. \emph{Annu.
		Rev. Phys. Chem.} \textbf{2022}, \emph{73}, 73--96\relax
	\mciteBstWouldAddEndPuncttrue
	\mciteSetBstMidEndSepPunct{\mcitedefaultmidpunct}
	{\mcitedefaultendpunct}{\mcitedefaultseppunct}\relax
	\EndOfBibitem
	\bibitem[Tobias \latin{et~al.}(2022)Tobias, Matsuda, Li, Miller, Carroll,
	Bilitewski, Rey, and Ye]{doi:10.1126/science.abn8525}
	Tobias,~W.~G.; Matsuda,~K.; Li,~J.-R.; Miller,~C.; Carroll,~A.~N.;
	Bilitewski,~T.; Rey,~A.~M.; Ye,~J. Reactions between layer-resolved molecules
	mediated by dipolar spin exchange. \emph{Science} \textbf{2022}, \emph{375},
	1299--1303\relax
	\mciteBstWouldAddEndPuncttrue
	\mciteSetBstMidEndSepPunct{\mcitedefaultmidpunct}
	{\mcitedefaultendpunct}{\mcitedefaultseppunct}\relax
	\EndOfBibitem
	\bibitem[Yang \latin{et~al.}(2022)Yang, Wang, Su, Cao, Zhang, Rui, Zhao, Bai,
	and Pan]{Yang2022}
	Yang,~H.; Wang,~X.~Y.; Su,~Z.; Cao,~J.; Zhang,~D.~C.; Rui,~J.; Zhao,~B.;
	Bai,~C.~L.; Pan,~J.~W. {Evidence for the association of triatomic molecules
		in ultracold $^{23}$Na$^{40}$K + $^{40}$K mixtures}. \emph{Nature}
	\textbf{2022}, \emph{602}, 229--233\relax
	\mciteBstWouldAddEndPuncttrue
	\mciteSetBstMidEndSepPunct{\mcitedefaultmidpunct}
	{\mcitedefaultendpunct}{\mcitedefaultseppunct}\relax
	\EndOfBibitem
	\bibitem[Jones \latin{et~al.}(2006)Jones, Tiesinga, Lett, and
	Julienne]{RevModPhys.78.483}
	Jones,~K.~M.; Tiesinga,~E.; Lett,~P.~D.; Julienne,~P.~S. Ultracold
	photoassociation spectroscopy: Long-range molecules and atomic scattering.
	\emph{Rev. Mod. Phys.} \textbf{2006}, \emph{78}, 483--535\relax
	\mciteBstWouldAddEndPuncttrue
	\mciteSetBstMidEndSepPunct{\mcitedefaultmidpunct}
	{\mcitedefaultendpunct}{\mcitedefaultseppunct}\relax
	\EndOfBibitem
	\bibitem[Yang \latin{et~al.}(2020)Yang, Huang, Hu, Xie, and
	Guo]{10.1063/5.0014805}
	Yang,~D.; Huang,~J.; Hu,~X.; Xie,~D.; Guo,~H. {Statistical quantum mechanical
		approach to diatom–diatom capture dynamics and application to ultracold KRb
		+ KRb reaction}. \emph{J. Chem. Phys.} \textbf{2020}, \emph{152},
	241103\relax
	\mciteBstWouldAddEndPuncttrue
	\mciteSetBstMidEndSepPunct{\mcitedefaultmidpunct}
	{\mcitedefaultendpunct}{\mcitedefaultseppunct}\relax
	\EndOfBibitem
	\bibitem[Huang \latin{et~al.}(2021)Huang, Yang, Zuo, Hu, Xie, and
	Guo]{doi:10.1021/acs.jpca.1c04506}
	Huang,~J.; Yang,~D.; Zuo,~J.; Hu,~X.; Xie,~D.; Guo,~H. Full-Dimensional Global
	Potential Energy Surface for the KRb + KRb $\rightarrow$ K$_{2}$Rb$_{2}$* $\rightarrow$ K$_{2}$ +
	Rb$_{2}$ Reaction with Accurate Long-Range Interactions and Quantum
	Statistical Calculation of the Product State Distribution under Ultracold
	Conditions. \emph{J. Phys. Chem. A} \textbf{2021}, \emph{125}, 6198--6206\relax
	\mciteBstWouldAddEndPuncttrue
	\mciteSetBstMidEndSepPunct{\mcitedefaultmidpunct}
	{\mcitedefaultendpunct}{\mcitedefaultseppunct}\relax
	\EndOfBibitem
	\bibitem[Buren(2023)]{D3CP01753B}
	Buren,~B. A neural network potential energy surface for the Li + LiNa $\rightarrow$
	Li$_{2}$ + Na reaction and quantum dynamics study from ultracold to thermal
	energies. \emph{Phys. Chem. Chem. Phys.} \textbf{2023}, \emph{25},
	19024--19036\relax
	\mciteBstWouldAddEndPuncttrue
	\mciteSetBstMidEndSepPunct{\mcitedefaultmidpunct}
	{\mcitedefaultendpunct}{\mcitedefaultseppunct}\relax
	\EndOfBibitem
	\bibitem[Blank \latin{et~al.}(1995)Blank, Brown, Calhoun, and
	Doren]{10.1063/1.469597}
	Blank,~T.~B.; Brown,~S.~D.; Calhoun,~A.~W.; Doren,~D.~J. {Neural network models
		of potential energy surfaces}. \emph{J. Chem. Phys.} \textbf{1995},
	\emph{103}, 4129--4137\relax
	\mciteBstWouldAddEndPuncttrue
	\mciteSetBstMidEndSepPunct{\mcitedefaultmidpunct}
	{\mcitedefaultendpunct}{\mcitedefaultseppunct}\relax
	\EndOfBibitem
	\bibitem[Brown \latin{et~al.}(1996)Brown, Gibbs, and Clary]{10.1063/1.472596}
     Brown,~D. F.~R.; Gibbs,~M.~N.; Clary,~D.~C. {Combining ab initio
		computations, neural networks, and diffusion Monte Carlo: An efficient method
		to treat weakly bound molecules}. \emph{J. Chem. Phys.} \textbf{1996},
	\emph{105}, 7597--7604\relax
	\mciteBstWouldAddEndPuncttrue
	\mciteSetBstMidEndSepPunct{\mcitedefaultmidpunct}
	{\mcitedefaultendpunct}{\mcitedefaultseppunct}\relax
	\EndOfBibitem
	\bibitem[Handley and Popelier(2010)Handley, and
	Popelier]{doi:10.1021/jp9105585}
	Handley,~C.~M.; Popelier,~P. L.~A. Potential Energy Surfaces Fitted by
	Artificial Neural Networks. \emph{J. Phys. Chem. A} \textbf{2010},
	\emph{114}, 3371--3383\relax
	\mciteBstWouldAddEndPuncttrue
	\mciteSetBstMidEndSepPunct{\mcitedefaultmidpunct}
	{\mcitedefaultendpunct}{\mcitedefaultseppunct}\relax
	\EndOfBibitem
	\bibitem[Behler(2011)]{C1CP21668F}
	Behler,~J. Neural network potential-energy surfaces in chemistry: a tool for
	large-scale simulations. \emph{Phys. Chem. Chem. Phys.} \textbf{2011},
	\emph{13}, 17930--17955\relax
	\mciteBstWouldAddEndPuncttrue
	\mciteSetBstMidEndSepPunct{\mcitedefaultmidpunct}
	{\mcitedefaultendpunct}{\mcitedefaultseppunct}\relax
	\EndOfBibitem
	\bibitem[Chen \latin{et~al.}(2013)Chen, Xu, Xu, and Zhang]{10.1063/1.4811109}
	Chen,~J.; Xu,~X.; Xu,~X.; Zhang,~D.~H. {Communication: An accurate global
		potential energy surface for the OH + CO $\rightarrow$ H + CO$_{2}$ reaction using
		neural networks}. \emph{J. Chem. Phys.} \textbf{2013}, \emph{138},
	221104\relax
	\mciteBstWouldAddEndPuncttrue
	\mciteSetBstMidEndSepPunct{\mcitedefaultmidpunct}
	{\mcitedefaultendpunct}{\mcitedefaultseppunct}\relax
	\EndOfBibitem
	\bibitem[Jiang and Guo(2013)Jiang, and Guo]{10.1063/1.4817187}
	Jiang,~B.; Guo,~H. {Permutation invariant polynomial neural network approach to
		fitting potential energy surfaces}. \emph{J. Chem. Phys.} \textbf{2013},
	\emph{139}, 054112\relax
	\mciteBstWouldAddEndPuncttrue
	\mciteSetBstMidEndSepPunct{\mcitedefaultmidpunct}
	{\mcitedefaultendpunct}{\mcitedefaultseppunct}\relax
	\EndOfBibitem
	\bibitem[Liu \latin{et~al.}(2016)Liu, Zhang, Fu, Yang, and Zhang]{C5SC03689E}
	Liu,~T.; Zhang,~Z.; Fu,~B.; Yang,~X.; Zhang,~D.~H. A seven-dimensional quantum
	dynamics study of the dissociative chemisorption of H$_{2}$O on Cu(111):
	effects of azimuthal angles and azimuthal angle-averaging. \emph{Chem. Sci.}
	\textbf{2016}, \emph{7}, 1840--1845\relax
	\mciteBstWouldAddEndPuncttrue
	\mciteSetBstMidEndSepPunct{\mcitedefaultmidpunct}
	{\mcitedefaultendpunct}{\mcitedefaultseppunct}\relax
	\EndOfBibitem
	\bibitem[Shao \latin{et~al.}(2016)Shao, Chen, Zhao, and
	Zhang]{10.1063/1.4961454}
	Shao,~K.; Chen,~J.; Zhao,~Z.; Zhang,~D.~H. {Communication: Fitting potential
		energy surfaces with fundamental invariant neural network}. \emph{J. Chem.
		Phys.} \textbf{2016}, \emph{145}, 071101\relax
	\mciteBstWouldAddEndPuncttrue
	\mciteSetBstMidEndSepPunct{\mcitedefaultmidpunct}
	{\mcitedefaultendpunct}{\mcitedefaultseppunct}\relax
	\EndOfBibitem
	\bibitem[Shakouri \latin{et~al.}(2017)Shakouri, Behler, Meyer, and
	Kroes]{doi:10.1021/acs.jpclett.7b00784}
	Shakouri,~K.; Behler,~J.; Meyer,~J.; Kroes,~G.-J. Accurate Neural Network
	Description of Surface Phonons in Reactive Gas–Surface Dynamics: N$_{2}$ +
	Ru(0001). \emph{J. Phys. Chem. Lett.} \textbf{2017}, \emph{8}, 2131--2136\relax
	\mciteBstWouldAddEndPuncttrue
	\mciteSetBstMidEndSepPunct{\mcitedefaultmidpunct}
	{\mcitedefaultendpunct}{\mcitedefaultseppunct}\relax
	\EndOfBibitem
	\bibitem[Anstine and Isayev(2023)Anstine, and
	Isayev]{doi:10.1021/acs.jpca.2c06778}
	Anstine,~D.~M.; Isayev,~O. Machine Learning Interatomic Potentials and
	Long-Range Physics. \emph{J. Phys. Chem. A} \textbf{2023}, \emph{127},
	2417--2431\relax
	\mciteBstWouldAddEndPuncttrue
	\mciteSetBstMidEndSepPunct{\mcitedefaultmidpunct}
	{\mcitedefaultendpunct}{\mcitedefaultseppunct}\relax
	\EndOfBibitem
	\bibitem[Tu \latin{et~al.}(2023)Tu, Rezajooei, Johnson, and Rowley]{D2DD00150K}
	Tu,~N. T.~P.; Rezajooei,~N.; Johnson,~E.~R.; Rowley,~C.~N. A neural network
	potential with rigorous treatment of long-range dispersion. \emph{Digital
		Discovery} \textbf{2023}, \emph{2}, 718--727\relax
	\mciteBstWouldAddEndPuncttrue
	\mciteSetBstMidEndSepPunct{\mcitedefaultmidpunct}
	{\mcitedefaultendpunct}{\mcitedefaultseppunct}\relax
	\EndOfBibitem
	\bibitem[Unke and Meuwly(2019)Unke, and Meuwly]{doi:10.1021/acs.jctc.9b00181}
	Unke,~O.~T.; Meuwly,~M. PhysNet: A Neural Network for Predicting Energies,
	Forces, Dipole Moments, and Partial Charges. \emph{J. Chem. Theory Comput.}
	\textbf{2019}, \emph{15}, 3678--3693\relax
	\mciteBstWouldAddEndPuncttrue
	\mciteSetBstMidEndSepPunct{\mcitedefaultmidpunct}
	{\mcitedefaultendpunct}{\mcitedefaultseppunct}\relax
	\EndOfBibitem
	\bibitem[Zhang \latin{et~al.}(2022)Zhang, Wang, Muniz, Panagiotopoulos, Car,
	and E]{10.1063/5.0083669}
	Zhang,~L.; Wang,~H.; Muniz,~M.~C.; Panagiotopoulos,~A.~Z.; Car,~R.; E,~W. {A
		deep potential model with long-range electrostatic interactions}. \emph{J.
		Chem. Phys.} \textbf{2022}, \emph{156}, 124107\relax
	\mciteBstWouldAddEndPuncttrue
	\mciteSetBstMidEndSepPunct{\mcitedefaultmidpunct}
	{\mcitedefaultendpunct}{\mcitedefaultseppunct}\relax
	\EndOfBibitem
	\bibitem[Li \latin{et~al.}(2023)Li, Zhai, and Li]{doi:10.1021/acs.jctc.2c01049}
	Li,~Y.; Zhai,~Y.; Li,~H. MLRNet: Combining the Physics-Motivated Potential
	Models with Neural Networks for Intermolecular Potential Energy Surface
	Construction. \emph{J. Chem. Theory Comput.} \textbf{2023}, \emph{19},
	1421--1431\relax
	\mciteBstWouldAddEndPuncttrue
	\mciteSetBstMidEndSepPunct{\mcitedefaultmidpunct}
	{\mcitedefaultendpunct}{\mcitedefaultseppunct}\relax
	\EndOfBibitem
	\bibitem[Werner \latin{et~al.}(2012)Werner, Knowles, Knizia, Manby, and
	Schütz]{molpro}
	Werner,~H.-J.; Knowles,~P.~J.; Knizia,~G.; Manby,~F.~R.; Schütz,~M. Molpro: a
	general-purpose quantum chemistry program package. \emph{WIREs Comput. Mol.
		Sci.} \textbf{2012}, \emph{2}, 242--253\relax
	\mciteBstWouldAddEndPuncttrue
	\mciteSetBstMidEndSepPunct{\mcitedefaultmidpunct}
	{\mcitedefaultendpunct}{\mcitedefaultseppunct}\relax
	\EndOfBibitem
	\bibitem[Deegan and Knowles(1994)Deegan, and Knowles]{DEEGAN1994321}
	Deegan,~M.~J.; Knowles,~P.~J. Perturbative corrections to account for triple
	excitations in closed and open shell coupled cluster theories. \emph{Chem.
		Phys. Lett.} \textbf{1994}, \emph{227}, 321--326\relax
	\mciteBstWouldAddEndPuncttrue
	\mciteSetBstMidEndSepPunct{\mcitedefaultmidpunct}
	{\mcitedefaultendpunct}{\mcitedefaultseppunct}\relax
	\EndOfBibitem
	\bibitem[Knowles \latin{et~al.}(2000)Knowles, Hampel, and
	Werner]{10.1063/1.480886}
	Knowles,~P.~J.; Hampel,~C.; Werner,~H.-J. {Erratum: “Coupled cluster theory
		for high spin, open shell reference wave functions” [ J. Chem. Phys. 99,
		5219 (1993)]}. \emph{J. Chem. Phys.} \textbf{2000}, \emph{112},
	3106--3107\relax
	\mciteBstWouldAddEndPuncttrue
	\mciteSetBstMidEndSepPunct{\mcitedefaultmidpunct}
	{\mcitedefaultendpunct}{\mcitedefaultseppunct}\relax
	\EndOfBibitem
	\bibitem[Dunning(1989)]{10.1063/1.456153}
	Dunning,~J.,~Thom~H. {Gaussian basis sets for use in correlated molecular
		calculations. I. The atoms boron through neon and hydrogen}. \emph{J. Chem.
		Phys.} \textbf{1989}, \emph{90}, 1007--1023\relax
	\mciteBstWouldAddEndPuncttrue
	\mciteSetBstMidEndSepPunct{\mcitedefaultmidpunct}
	{\mcitedefaultendpunct}{\mcitedefaultseppunct}\relax
	\EndOfBibitem
	\bibitem[Prascher \latin{et~al.}(2011)Prascher, Woon, Peterson, Dunning, and
	Wilson]{prascher2011gaussian}
	Prascher,~B.~P.; Woon,~D.~E.; Peterson,~K.~A.; Dunning,~T.~H.; Wilson,~A.~K.
	Gaussian basis sets for use in correlated molecular calculations. VII.
	Valence, core-valence, and scalar relativistic basis sets for Li, Be, Na, and
	Mg. \emph{Theor. Chem. Acc.} \textbf{2011}, \emph{128}, 69--82\relax
	\mciteBstWouldAddEndPuncttrue
	\mciteSetBstMidEndSepPunct{\mcitedefaultmidpunct}
	{\mcitedefaultendpunct}{\mcitedefaultseppunct}\relax
	\EndOfBibitem
	\bibitem[Vogels \latin{et~al.}(2018)Vogels, Karman, K{\l}os, Besemer, Onvlee,
	{Van Der Avoird}, Groenenboom, and {Van De Meerakker}]{Vogels2018}
	Vogels,~S.~N.; Karman,~T.; K{\l}os,~J.; Besemer,~M.; Onvlee,~J.; {Van Der
		Avoird},~A.; Groenenboom,~G.~C.; {Van De Meerakker},~S.~Y. {Scattering
		resonances in bimolecular collisions between NO radicals and H$_{2}$
		challenge the theoretical gold standard}. \emph{Nat. Chem.} \textbf{2018},
	\emph{10}, 435--440\relax
	\mciteBstWouldAddEndPuncttrue
	\mciteSetBstMidEndSepPunct{\mcitedefaultmidpunct}
	{\mcitedefaultendpunct}{\mcitedefaultseppunct}\relax
	\EndOfBibitem
\end{mcitethebibliography}
\providecommand{\latin}[1]{#1}
\makeatletter
\providecommand{\doi}
{\begingroup\let\do\@makeother\dospecials
	\catcode`\{=1 \catcode`\}=2 \doi@aux}
\providecommand{\doi@aux}[1]{\endgroup\texttt{#1}}
\makeatother
\providecommand*\mcitethebibliography{\thebibliography}
\csname @ifundefined\endcsname{endmcitethebibliography}
{\let\endmcitethebibliography\endthebibliography}{}

\end{document}